\documentclass[twocolumn,aps,pr,superscriptaddress,showpacs]{revtex4}
\usepackage{amsmath,bm}%
\usepackage{graphicx}%

\begin{document}

\draft
\title{  Isoscaling in the Lattice Gas Model}

\author{Y. G. Ma}
\thanks{Email: ygma@sinr.ac.cn}
\affiliation{Shanghai Institute of Applied Physics, Chinese
Academy of Sciences, P.O. Box 800-204, Shanghai 201800, China}
\affiliation{China Center of Advanced Sciences and Technology
(World Laboratory), P.O. Box 8730, Beijing 100080, China}
\author{K. Wang}
\author{Y. B. Wei}
\author {G. L. Ma}
\affiliation{Shanghai Institute of Applied Physics, Chinese
Academy of Sciences, P.O. Box 800-204, Shanghai 201800, China}
\author{X. Z. Cai}
\author{J. G. Chen}
\author{D. Q. Fang}
\author{W. Guo}
\affiliation{Shanghai Institute of Applied Physics, Chinese
Academy of Sciences, P.O. Box 800-204, Shanghai 201800, China}
\author{W. Q. Shen}
\affiliation{Shanghai Institute of Applied Physics, Chinese
Academy of Sciences, P.O. Box 800-204, Shanghai 201800, China}
\affiliation{China Center of Advanced Sciences and Technology
(World Laboratory), P.O. Box 8730, Beijing 100080, China}
\author{W. D. Tian}
\author{C. Zhong}
\affiliation{Shanghai Institute of Applied Physics, Chinese
Academy of Sciences, P.O. Box 800-204, Shanghai 201800, China}


\date{\today}

\nopagebreak

\begin{abstract}
The isoscaling behavior is investigated using the
isotopic/isobaric yields from the equilibrated thermal source
which is prepared by the lattice gas model (LGM)  for lighter
systems with  A = 36. The isoscaling parameters $\alpha$ and
-$\beta$ are observed to  drop with temperature for the LGM with
the asymmetric nucleon-nucleon potential. However, the isoscaling
parameters do not show temperature dependence for the LGM with the
symmetric nucleon-nucleon potential. The relative neutron or
proton density shows a nearly linear relation with the N/Z (
neutron to proton ratio ) of system.

\end{abstract}



\maketitle

Isoscaling has been observed in a variety of reactions under the
conditions of statistical emission and  equal temperature recently
by Tsang et al \cite{TsangPRL,Tsang2,Tsang3}. This kind of scaling
means that the ratio R$_{21}$(N,Z) of the yields of a given
fragment (N,Z) exhibits an exponential dependence on N and Z when
these fragments are produced in two reactions with different
isospin asymmetry, but at the same temperature. Experimentally the
isoscaling has been explored in
 various reaction mechanisms, ranging from the evaporation \cite{TsangPRL},
fission \cite{Friedman,Veselsky2} and deep inelastic reaction at
low energies to the projectile fragmentation \cite{TAMU1,Veselsky}
and multi-fragmentation at intermediate energy
\cite{TsangPRL,LiuTX,Geraci}. While, the isoscaling has been
extensively examined in different theoretical frameworks, ranging
from dynamical model, such as Anti-symmetrical Molecular Dynamics
model \cite{Ono} and BUU model \cite{LiuTX}, to statistical
models, such as Expansion Emission Source Model and statistical
multi-fragmentation model \cite{Tsang2,Tsang3,Botvina,Souza}. From
all these reaction mechanisms and models, it looks that isoscaling
is a robust probe to relate with the symmetrical term of the
nuclear equation of state.

Typically, the investigations of isoscaling  focused on yields of
light fragments with Z=2-8 originating from de-excitation of
massive hot systems produced using reactions of mass symmetric
projectile and target at intermediate energies, such as
$^{112,124}Sn$ +  $^{112,124}Sn$  in Michigan State University
(MSU) data \cite{TsangPRL,Tsang2,Tsang3} or by reactions of
high-energy light particle with massive target nucleus
\cite{Botvina,Russia}. In a recent article \cite{TAMU1}, the
isoscaling using the heavy projectile residue from the reactions
of 25 MeV/nucleon $^{86}$Kr projectiles with $^{124}$Sn,$^{112}$Sn
and $^{64}$Ni, $^{58}$Ni targets which was performed at Texas A\&M
University (TAMU) and  the isoscaling phenomenon on the full
sample of fragments emitted by the hot thermally equilibrated
quasi-projectiles with mass A = 20-30 are also reported
\cite{Veselsky}.

In this study, we present an isoscaling analysis for the light
fragments from thermal sources which are produced by the lattice
gas model. Instead of the fixed charge number of the reaction
system in MSU data or TAMU data, here we fixed the source mass for
lighter system, but changing the charge number and neutron number.
The isospin fractionation is also observed via the relative free
neutron and free proton density which is obtained by the
isoscaling parameters.

A thermally equilibrated system undergoing statistical decay can be,
within grand-canonical approach, characterized by a yield of fragments
with neutron and proton numbers N and Z
\cite{Albergo,Randrup}:
\begin{equation}
        Y(N,Z) = F(N,Z)\exp\frac{B(N,Z)}{T}\exp(\frac{N \mu_{n}}{T} + \frac{Z \mu_{p}}{T})
\label{eqn1}
\end{equation}
where $F(N,Z)$ represents contribution due to the secondary decay from
particle stable and  unstable states to the ground state; $\mu_{n}$
and $\mu_{p}$ are the free neutron and proton chemical potentials; $B(N,Z)$ is
the ground state binding energy of the fragment, and $T$ is the
temperature.

The ratio of the isotope yields from two different systems, having
similar excitation energies and similar masses, but differing only
in $N/Z$, cancels out the effect of secondary decay and provides
information about the excited primary fragments \cite{TsangPRL}.
Within the grand-canonical approximation ( Eq.(\ref{eqn1}) ), the
ratio $Y_{2}(N,Z)/Y_{1}(N,Z)$ assumes the form
\begin{equation}
     R_{21}(N,Z) = Y_{2}(N,Z)/Y_{1}(N,Z) = C \exp(\alpha N  + \beta Z)
\end{equation}
with  $\alpha$ = $\Delta \mu_{n}/T$
and $\beta$ = $\Delta \mu_{p}$/T, with $\Delta \mu_{n}$ and $\Delta \mu_{p}$
being the differences in the free neutron  and proton chemical potentials
of the fragmenting systems. $C$ is an  overall normalization constant.

The tool we will use here is the isospin dependent lattice gas
model (LGM). The lattice gas model was developed to describe the
liquid-gas phase transition for atomic system by Lee and
 Yang \cite{Yang52}. The same model has already been applied to
nuclear physics for isospin symmetrical systems in the
grand-canonical ensemble \cite{Biro86} with a sampling of the
canonical ensemble
\cite{Camp97,Jpan96,Mull97,Jpan95,Jpan98,Gulm98,Ma99}, and also
for isospin asymmetrical nuclear matter in the mean field
approximation \cite{Sray97}.  Here we will make a brief
description for the models.

In the lattice gas  model, $A$ (= $N + Z$) nucleons with an
occupation number $s$ which is defined $s$ = 1 (-1) for a proton
(neutron) or $s$ = 0 for a vacancy, are placed on the $L$ sites of
lattice. Nucleons in the nearest neighboring sites
interact with an energy $\epsilon_{s_i s_j}$. The hamiltonian
is written as
\begin{equation}
E = \sum_{i=1}^{A} \frac{P_i^2}{2m} -
\sum_{i < j} \epsilon_{s_i s_j}s_i s_j .
\end{equation}
In order to investigate the symmetrical term of nuclear potential
in this model, we use two sets of parameters: one is an attractive
potential constant  $\epsilon_{s_i s_j}$ between the neutron and
protons but no interaction between like nucleon, i.e. proton and
proton or neutron and neutron, namely
\begin{eqnarray}
 \epsilon_{nn} \ &=&\ \epsilon_{pp} \ = \ 0. MeV \nonumber , \\
 \epsilon_{pn} \ &=&\ - 5.33 MeV.
\end{eqnarray}
This potential results in an asymmetrical potential among
different kind of nucleons, hence it is an isospin dependent
potential. For simplicity we call the calculation with this
potential as isoLGM thereafter. Another set is the same
interaction constant between like nucleons or unlike nucleons,
i.e.
\begin{equation}
 \epsilon_{pn} = \epsilon_{nn} = \epsilon_{pp} = - 5.33 MeV.
\end{equation}
In this case, the nucleon potential is symmetrical among all
nucleons, i.e. isospin independent potential. For simplicity, we
call the calculation with Eq.(5) as noisoLGM thereafter. In this
work, mostly we use isoLGM to explore isoscaling behavior, but we
will also use noisoLGM to compare  the isoscaling results.

In the LGM simulation, a three-dimension cubic lattice with $L$ sites is used. The
freeze-out density of disassembling system is assumed to be
 $\rho_f$ = $\frac{A}{L} \rho_0$, where $\rho_0$ is the normal
 nuclear density. The disassembly of the system
is to be calculated at $\rho_f$, beyond which nucleons are too far
apart to interact.  Nucleons are put into lattice by Monte Carlo
Metropolis sampling. Once the nucleons have been placed we also
ascribe to each of them a momentum by Monte Carlo samplings of
Maxwell-Boltzmann distribution.

Once this is done the LGM immediately gives the cluster
distribution using the rule that two nucleons are part of the
same cluster if $P_r^2/2\mu - \epsilon_{s_i s_j}s_i
s_j < 0 $. This method is similar to the Coniglio-Klein's
prescription \cite{Coni80} in condensed matter physics.

In this paper we choose the small size nuclei with $A$ = 36 as
emission sources. Four isotonic sources, namely $^{36}Ca$,
$^{36}Ar$, $^{36}S$ and $^{36}Si$, corresponding to N/Z = 0.8,
1.0, 1.25 and 1.57, respectively, are simulated. In all cases, the
freeze-out density $\rho_f$ is chosen to be 0.563 $\rho_0$, which
corresponds to $4^3$ cubic lattice is used. 10000 events were
simulated for each $T$ which ensures good statistics for results.

\begin{figure}
\vspace{-0.3truein}
\includegraphics[scale=0.4]{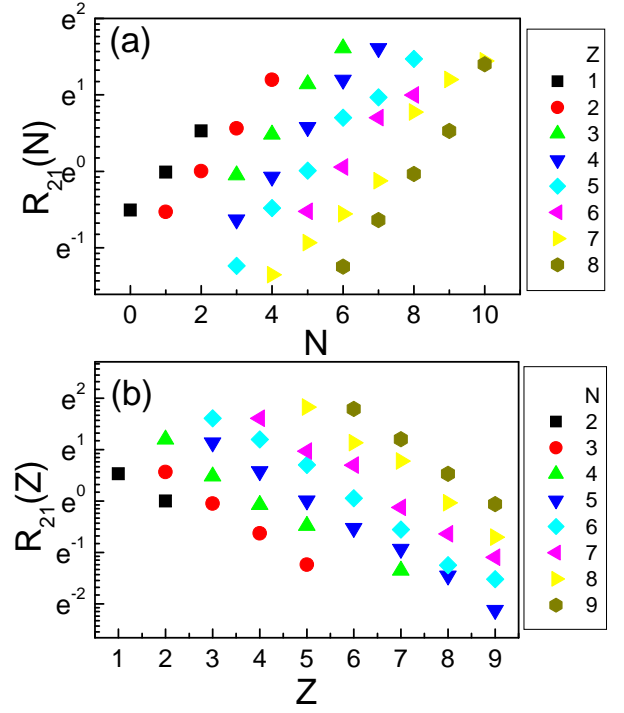}
\vspace{-0.2truein} \caption{\footnotesize (Color online)
Isoscaling behavior of $R_{21}(N)$ (the upper panel) and
$R_{21}(Z)$ (the lower panel) at $T$ = 5.0 MeV.} \label{scaling}
\end{figure}

\begin{figure}
\vspace{-1.truein}
\includegraphics[scale=0.4]{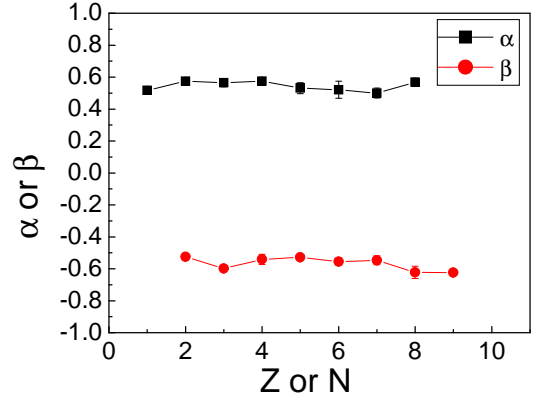}
\vspace{-1.1truein} \caption{\footnotesize (Color online)
Isoscaling parameter $\alpha$ and $\beta$ extracted in fixed Z
from Fig.~\ref{scaling}(a) and N from Fig.~\ref{scaling}(b),
respectively.} \label{para_N}
\end{figure}

As an example, in Fig. \ref{scaling}, we present the isotopic
scaling (the upper panel) and the isobaric scaling (the lower
panel) from the isospin dependent lattice gas model for $^{36}S$
to $^{36}Ca$, respectively, at $T$ = 5.0 MeV. The  charged
particles with Z$\leq$8 and 2 $\leq$ N $\leq$ 9 have been
accumulated to perform the ratios. There exist good linear
behavior in semi-log plot. In order to extract the isoscaling
parameters $\alpha$ and $\beta$, we use $R_{21}(N) = C exp(\alpha
N)$ to obtain $\alpha$ for a given $Z$ (Fig. \ref{scaling}(a)) and
use $R_{21}(Z) = C exp(\beta Z)$ to obtain $\beta$ for a given $N$
(Fig. \ref{scaling}(b)), respectively. Fig.~\ref{para_N} shows the
extracted $\alpha$ or $\beta$ versus
 $Z$ or $N$. Obviously, $\alpha$ or $\beta$ keeps the same value in the
wide $Z$ or $N$ range and their absolute values are almost the
same which is due to the absence of Coulomb interaction in the
lattice gas model.

\begin{figure}
\vspace{-1.truein}
\includegraphics[scale=0.4]{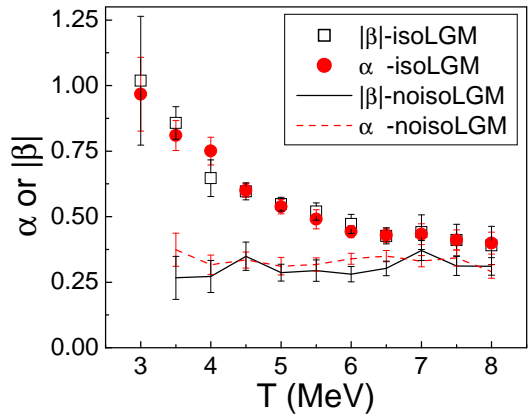}
\vspace{-0.8truein} \caption{\footnotesize (Color online) The
absolute value of the isoscaling parameters $\alpha$ and $\beta$
as a function of $T$ from the yield ratios from the sources of
$^{36}S$ and $^{36}Ca$. The scattering points are the calculation
results for LGM with asymmetrical nucleon-nucleon potential
(Eq.(4)) and the lines represent the results for the LGM with the
same nucleon-nucleon potential for unlike and like nucleon pairs
(Eq.(5)). See text for details.} \label{para}
\end{figure}

\begin{figure}
\includegraphics[scale=0.4]{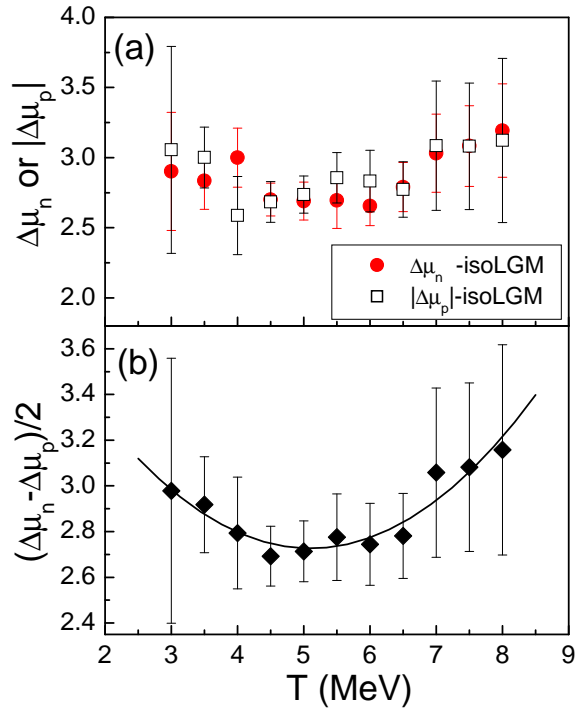}
\caption{\footnotesize (a) (Color online) The absolute value of
the difference of neutron and proton chemical potential ($\Delta
\mu _n$ and $|\Delta \mu _p|$) as a function of T.  (b) ($\Delta
\mu_n - \Delta \mu_p$)/2 as a function of temperature. Solid
points are calculated results and line is its second polynomial
fit. See text for details.} \label{chemical}
\end{figure}

Fig.~\ref{para} shows the temperature dependence of the absolute
values of isoscaling parameters $\alpha$ and $\beta$. Here the
scattering points are the calculation results for LGM with
asymmetrical nucleon-nucleon potential and the lines represent the
results for the LGM with the same nucleon-nucleon potential for
unlike and like nucleon pairs. A decreasing trend of the values is
clearly seen when the asymmetrical nucleon-nucleon potential is
used in LGM, which indicates that the isospin dependence of the
fragment yields becomes weak with increasing temperature.
Considering that $\alpha$ = $\Delta \mu_{n}/T$ and $\beta$ =
$\Delta \mu_{p}$/T, we can deduce the differences in free neutron
and proton chemical potentials of the fragmenting systems which is
shown in Fig.~\ref{chemical}(a). Within the error bars, it looks
like that $\Delta \mu_{n}$ and  $\Delta \mu_{p}$ keep constant in
the case of asymmetrical potential (Eq.(4)).
However, a slight and wide valley can be also identified around
$T$ = 5 MeV as well as a slight kink shows in SMM model for Sn
systems \cite{Botvina}. For our system, this valley may be related
to the liquid gas phase change for the similar system around 5 MeV
in the same model calculation as well as the data \cite{Ma2004}.
This valley becomes more obvious if we plot the values of ($\Delta
\mu_n - \Delta \mu_p$)/2 as a function of temperature as suggested
in Ref.~\cite{Botvina,Veselsky}. 
Fig.~\ref{chemical}(b) shows that a turning point seems to occur
around 5 MeV. Of course, the error bars look larger for lower and
higher temperatures, because of the poor statistics for diverse
cluster species due to the evaporation mechanism in low $T$ and
the vaporization mechanism in higher $T$. 
Around this point, an apparent critical behavior has been observed
in the disassembly of hot nuclei with A$\sim$36 in TAMU-NIMROD
experimental data and model calculations \cite{Ma2004} by a wide
variety of observables: such as the maximal fluctuations, critical
exponent analysis and fragment topological structure, namely
nuclear Zipf law of Ma \cite{MaPRL} and the heaviest and second
heaviest correlation etc. Hence, this turning point of the
difference of neutron and proton chemical potential might give an
additional evidence of the chemical phase separation when the
liquid gas phase transition occurs. Experimentally, this kind of
turning point has been observed recently for the quasi-projectile
from the peripheral collisions of $^{28}Si$ + $^{112,124}Sn$ at 30
and 50 MeV/nucleon in another TAMU data and can be understood as a
signal of the onset of separation into isospin asymmetric dilute
and isospin symmetric dense phase in a recent paper by Veselsky et
al \cite{Veselsky}.

\begin{figure}
\vspace{-0.2truein}
\includegraphics[scale=0.4]{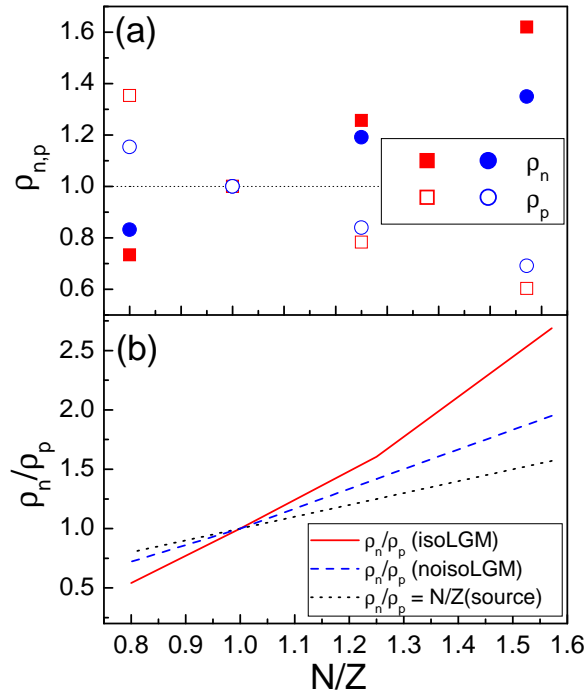}
\vspace{-0.1truein} \caption{\footnotesize (Color online) (a) The
relative free neutron (solid symbols) or proton (open symbols)
density as a function of the N/Z of system at T = 5 MeV for the
isoLGM case (squares) and noisoLGM case (circles). (b) The solid
and dashed line represents the ratio $\rho_n/\rho_p$  for isoLGM
case and noisoLGM case, respectively. The dotted line shows the
initial $\rho_n/\rho_p$ assuming neutrons and protons
homogeneously distributed in a volume proportional to the nucleon
number. } \label{rho}
\end{figure}

Further, we shall investigate the effect of the asymmetrical
nucleon-nucleon potential in the lattice gas model on the
isoscalig behavior. We take the same potential between like
nucleons and  unlike nucleons in the lattice gas model (Eq.(5)) to
compare with LGM with isospin asymmetrical nucleon-nucleon
potential (Eq.(4)). In this case, we observe that temperature
dependence of $\alpha$ and $\beta$ vanishes and their absolute
values are smaller than the isoLGM case as shown by the lines in
Fig.~\ref{para}. The insensitivity of $\alpha$ and $\beta$ to
temperature means that the isoscaling of fragment yields only
stems from the initial N/Z of the system. This is possible since
the nucleon-nucleon potential is the same for like nucleons and
unlike nucleons in this noisoLGM case. When temperature increases,
the yield of light clusters also increase, but their isospin
population does not change between the different cluster species,
which results that no dependence for the isotopic or isobaric
ratio on temperature. In this case, the system looks like a single
component fluid, the chemical phase separation can  not occur.
However, for the isoLGM, the nucleon-nucleon potential is
asymmetrical between like pair and unlike pair, it will give  an
additional isoscaling contribution due to the isospin dependent
equation of state except from the trivial contribution from the
initial difference of the isospin of the source, which results in
a bigger absolute value of $\alpha$ and  $\beta$ than the noisoLGM
case. In this context, the information on the equation of state of
the asymmetric nucleon-nucleon potential could be extracted by the
temperature dependence of isoscaling parameters.

Finally, we test the relative free neutron density and proton
density, which can be deduced by the equations:
\begin{eqnarray}
\rho_n \ &=&\  exp(\alpha^{\prime} ),\\
 \rho_p \ &=&\  exp(\beta^{\prime} ).
\end{eqnarray}
in the lattice gas model.  Here $\alpha^{\prime}$  and
$\beta^{\prime}$ is the scaling parameters where we take
$Y_{1}(N,Z)$ from $^{36}Ar$ instead of $^{36}Ca$ as shown above.
Principally, the association of a number density with the
$exp(\Delta \mu/T)$ only is valid for a classical gas of free
particles. The connection in this context of the LGM where the
particles are interacting is not clear.
 However, as an attempt, we will still use the above equation to deduce the
 the relative  free neutron density and proton density.
  In Fig.~\ref{rho}(a) we show the
$\rho_n$ (solid symbols) and $\rho_p$ (open symbols) as a function
of N/Z of the systems at $T$ = 5 MeV for the isoLGM case (squares)
and the noisoLGM case (circles). In both cases, nearly linear
relations of $\rho_n$ and $\rho_p$ have been  observed with the
increasing of N/Z. The Fig.~\ref{rho}(b) shows the ratio
$\rho_n/\rho_p$. The solid (dashed) line represents the deduced
$\rho_n/\rho_p$ from solid and open  squares (circles) of
Fig.~\ref{rho}(a) with isoLGM (noisoLGM) calculation and the
dotted line is the initial value of $\rho_n/\rho_p$ which is
calculated assuming neutrons and protons homogeneously distributed
in a volume proportional to the nucleon number. In isoLGM case,
the values of $\rho_n/\rho_p$ for neutron-rich nuclei ($^{36}S$
and $^{36}Si$) is much larger than the initial value and while the
values of $\rho_n/\rho_p$ for proton-rich nuclei ($^{36}Ca$) is
much less than the initial value. While, in noisoLGM case,  the
values of $\rho_n/\rho_p$ also increases with the initial $N/Z$
value of sources, which is basically originated from the initial
difference of isospin of hot emitters. The stronger enhancement of
$\rho_n/\rho_p$ in the isoLGM case may indicate of a neutron
enrichment while a proton depletion in the nuclear gas phase. In
this context, it may be consistent with the isospin fractionation
effect which is a signal expected for the liquid gas phase
transition in asymmetrical systems
\cite{Geraci,Muller,Chomaz,Xu,Ma2000}. However, this
interpretation is not unique, since the larger $\rho_n/\rho_p$ can
be also directly attributable to the interaction and thus probably
not an increase in neutron density.

In summary, the isoscaling is investigated using the fragment
yield from the equilibrated thermal source with the same mass
number but different N/Z which was prepared by the lattice gas
model with the asymmetrical potential between like nucleons and
unlike nucleons. The isotopic scaling and isobaric scaling are
observed for light clusters and the isoscaling parameters $\alpha$
and $\beta$ are extracted from the ratios of $R_{21}(N,Z)$ for
fixed proton number or neutron number. It is found that $\alpha$
and -$\beta$ is almost the same and they drop with the
temperature. However, the difference of the neutron ($\Delta
\mu_n$) and proton ($\Delta \mu_p$) chemical potential does not
change much with temperature and a slight and wide valley for
($\Delta \mu_n - \Delta \mu_p$) is observed around T = 5 MeV even
though we have larger error bars for lower and higher
temperatures, may indicate the onset of phase separation into
isospin asymmetric dilute and isospin symmetric dense phase where
the liquid gas phase change occurs. The relative free neutron
density or proton density is attempted to be deduced from the
isoscaling parameter and they reveal a nearly linear relation to
the N/Z of the initial system. However, the values for
neutron-rich source are much larger than the initial value of N/Z
as well as the values for neutron-poor source are much less than
the initial value of N/Z. This may be from the isospin
fractionation effect, or can be directly attributable to the
interaction.

In addition, in order to investigate the effect of the
asymmetrical nucleon-nucleon potential in LGM, we also adopt the
same potential between like nucleons and  unlike nucleons in LGM.
In this case, isoscaling still remains but  the temperature
dependence of isoscaling parameter ($\alpha$ and $\beta$) vanishes
and their absolute values decrease. The insensitivity of the
isoscaling parameter to temperature means that the isospin
partition between different fragments is almost the same
regardless the excitation extent of system and the chemical phase
separation is absent in this case. However, for the isoLGM, the
asymmetrical nucleon-nucleon potential between like pair and
unlike pair gives an additional contribution from isospin
dependent equation of state to the isoscaling behavior except from
the trivial contribution from the initial difference of the
isospin of the source, which results in a bigger absolute value of
$\alpha$ and $\beta$ than the noisoLGM case. In this context,
information on the equation of state of the asymmetric
nucleon-nucleon potential could be extracted by the temperature
dependence of isoscaling parameters.

The authors appreciate Dr. Subal Das Gupta and Dr. Jicai Pan for
providing the original code of LGM and Dr. Betty Tsang and Dr.
Martin Veselsky for communications and discussions. This work was
supported partly by the Major State Basic Research Development
Program under Contract No G200077404, the National Natural Science
Foundation of China under (NNSFC) Grant No 10328259 and 10135030,
the Chinese Academy Sciences Grant for Distinguished Young
Scholars of NNSFC under Grant No 19725521 and the Shanghai
Phosphor Program Under Contract Number 03 QA 14066.

{}


\footnotesize
\begin{thebibliography}{}

\bibitem{TsangPRL} M. B. Tsang et al., Phys. Rev. Lett. {\bf 86}, 5023 (2001).
\bibitem{Tsang2} M. B. Tsang et al., Phys. Rev.  C {\bf 64}, 041603 (2001).
\bibitem{Tsang3} M. B. Tsang et al., Phys. Rev. C {\bf 64}, 054615 (2001).


\bibitem{Friedman}W. A. Friedman, Phys. Rev. C {\bf 69}, 031601(R)
(2004).

\bibitem{Veselsky2} M. Veselsky, G. A. Souliotis, and M. Jandel,
Phys. Rev. C {\bf 69}, 044607 (2004).

\bibitem{TAMU1}G. A. Souliotis et al., Phys. Rev. C {\bf 68}, 024605 (2003).
\bibitem{Veselsky} M. Veselsky et al., Phys. Rev. C {\bf 69}, 031602(R) (2004).
\bibitem{LiuTX}T. X. Liu et al., Phys. Rev. C {\bf 69}, 014603 (2004).
\bibitem{Geraci}E. Geraci et al., Nucl. Phys. A {\bf 732}, 173 (2004).

\bibitem{Ono}A. Ono et al., Phys. Rev. C {\bf 68}, 051601 (2003).

\bibitem{Botvina}A. S. Botvina, O. V. Lozhkin, and W. Trautmann, Phys. Rev. C {\bf 65}, 044610 (2002).

\bibitem{Souza}S. R. Souza et al.,  Phys. Rev. C {\bf 69}, 031607(R) (2004).

\bibitem{Russia}M. N. Andronenko, L. N. Andronenko, and W. Neubert, Prog. The. Phys. Suppl. {\bf 146}, 538 (2002).

\bibitem{Albergo}S. Albergo et al., Nuovo Cimento A {\bf 89}, 1 (1985).
\bibitem{Randrup}J. Randrup and S. E. Koonin, Nucl. Phys. A {\bf 356}, 223 (1981).

\bibitem{Yang52}T. D. Lee and C.N. Yang, Phys.\ Rev.\ {\bf 87}, 410 (1952).

\bibitem{Biro86}T. S. Biro {\it et\ al}., Nucl.\ Phys.\ A {\bf 459}, 692  (1986);
S.K. Samaddar  and J. Richert, Phys.\ Lett.\ B {\bf 218}, 381 (1989); Z.\ Phys.\ A {\bf 332}, 443 (1989);
J.M. Carmona {\it et\ al}., Nucl.\ Phys.\ A {\bf 643}, 115 (1998).

\bibitem{Mull97}W. F. J. M\"uller, Phys.\ Rev.\ C {\bf 56}, 2873 (1997).

\bibitem{Camp97}X. Campi  and H. Krivine, Nucl.\ Phys.\ A {\bf 620}, 46 (1997).

\bibitem{Jpan96}J. Pan  and S. Das Gupta, Phys.\ Rev.\ C {\bf 53}, 1319 (1996).

\bibitem{Jpan95}J. Pan  and S. Das Gupta, Phys.\ Lett.\ B {\bf 344}, 29 (1995);
 Phys.\ Rev.\ C {\bf 51}, 1384 (1995); Phys.\ Rev.\ Lett.\ {\bf 80}, 1182 (1998);
 S. Das Gupta {\it et\ al}.,  Nucl.\ Phys.\ A {\bf 621}, 897 (1997).

\bibitem{Jpan98} J. Pan and S. Das Gupta, Phys.\ Rev.\ C {\bf 57}, 1839 (1998).

\bibitem{Gulm98}F. Gullminelli and P. Chomaz, Phys.\ Rev.\ Lett.\ {\bf 82}, 1402 (1999).

\bibitem{Ma99}Y. G. Ma {\it et\ al}., Chin. Phys. Lett. {\bf 16}, 256 (1999);
Phys. Rev. C {\bf 60}, 24607 (1999).

\bibitem{Sray97}S. Ray {\it et\ al}., Phys.\ Lett.\ B {\bf 392}, 7 (1997).

\bibitem{Coni80}A. Coniglio and E. Klein, J.\ Phys.\ A {\bf 13}, 2775 (1980).

\bibitem{Ma2004}Y. G. Ma et al., Phys. Rev. C {\bf 69}, 031604(R)(2004)
and a long paper  to be submitted.

\bibitem{MaPRL} Y. G. Ma, Phys. Rev. Lett. {\bf 83}, 3617 (1999);
Eur. Phys. J. A {\bf 6}, 367 (1999).

\bibitem{Muller}H. Muller, B. D. Serot, Phys. Rev. C {\bf 52}, 2072 (1995).
\bibitem{Chomaz}Ph. Chomaz, F. Gulminelli, Phys. Lett. B {\bf 447}, 221 (1999).
\bibitem{Xu}H. S. Xu et al., Phys. Rev. Lett. {\bf 85}, 716 (2000).
\bibitem{Ma2000}Y. G. Ma, Acta. Phys. Sinica {\bf 49}, 654 (2000).
\end{thebibliography}
\end{document}